\documentclass[twocolumn]{aastex63}

\graphicspath{{./}{figures/}}
\usepackage{makecell}

\shorttitle{TDE Hosts Are Green \& Concentrated}
\shortauthors{Hammerstein et al.}

\begin{document}

\title{TDE Hosts are Green and Centrally Concentrated: Signatures of a Post-Merger System}

\author[0000-0002-5698-8703]{Erica Hammerstein}
\affil{Department of Astronomy, University of Maryland, College Park, MD 20742, USA}

\author{Suvi Gezari}
\affiliation{Department of Astronomy, University of Maryland, College Park, MD  20742, USA}
\affiliation{Joint Space-Science Institute, University of Maryland, College Park, MD 20742, USA}

\author{Sjoert van Velzen}
\affiliation{Department of Astronomy, University of Maryland, College Park, MD 20742, USA}
\affiliation{Center for Cosmology and Particle Physics, New York University, NY 10003}

\author[0000-0003-1673-970X]{S.~Bradley Cenko}
\affiliation{Astrophysics Science Division, NASA Goddard Space Flight Center, MC 661, Greenbelt, MD 20771, USA}
\affiliation{Joint Space-Science Institute, University of Maryland, College Park, MD 20742, USA}

\author{Nathaniel Roth}
\affil{Department of Astronomy, University of Maryland, College Park, MD 20742, USA}
\affil{Joint Space-Science Institute, University of Maryland, College Park, MD 20742, USA}

\author{Charlotte Ward}
\affil{Department of Astronomy, University of Maryland, College Park, MD 20742, USA}

\author{Sara Frederick}
\affil{Department of Astronomy, University of Maryland, College Park, MD 20742, USA}

\author{Tiara Hung}
\affil{Department of Astronomy and Astrophysics, University of California, Santa Cruz, CA 95064, USA}

\author{Matthew Graham}
\affiliation{Division of Physics, Mathematics, and Astronomy, California Institute of Technology, Pasadena, CA 91125, USA}

\author[0000-0002-2445-5275]{Ryan~J.~Foley}
\affiliation{Department of Astronomy and Astrophysics, University of California, Santa Cruz, CA 95064, USA}


\author[0000-0001-8018-5348]{Eric C. Bellm}
\affiliation{DIRAC Institute, Department of Astronomy, University of Washington, 3910 15th Avenue NE, Seattle, WA 98195, USA}

\author{Christopher Cannella}
\affiliation{Department of Electrical and Computer Engineering, Duke University, Durham, NC 27708, USA}

\author{Andrew J. Drake}
\affiliation{Division of Physics, Mathematics, and Astronomy, California Institute of Technology, Pasadena, CA 91125, USA}

\author[0000-0002-6540-1484]{Thomas Kupfer}
\affil{Texas Tech University, Department of Physics \& Astronomy, Box 41051, 79409, Lubbock, TX, USA}

\author[0000-0003-2451-5482]{Russ R. Laher}
\affiliation{IPAC, California Institute of Technology, 1200 E. California
             Blvd, Pasadena, CA 91125, USA}

\author[0000-0003-2242-0244]{Ashish~A.~Mahabal}
\affiliation{Division of Physics, Mathematics, and Astronomy, California Institute of Technology, Pasadena, CA 91125, USA}
\affiliation{Center for Data Driven Discovery, California Institute of Technology, Pasadena, CA 91125, USA}

\author[0000-0002-8532-9395]{Frank J. Masci}
\affiliation{IPAC, California Institute of Technology, 1200 E. California
             Blvd, Pasadena, CA 91125, USA}
             
\author{Reed Riddle}
\affiliation{Caltech Optical Observatories, California Institute of Technology, Pasadena, CA, USA}

\author{C\'esar Rojas-Bravo}
\affil{Department of Astronomy and Astrophysics, University of California, Santa Cruz, CA 95064, USA}

\author[0000-0001-7062-9726]{Roger Smith}
\affil{Caltech Optical Observatories, California Institute of Technology, Pasadena, CA, USA}

\correspondingauthor{Erica Hammerstein}
\email{ekhammer@astro.umd.edu}

\begin{abstract}
We study the properties of the galaxies hosting the first 19 tidal disruption events (TDEs) detected with the Zwicky Transient Facility (ZTF) within the context of a carefully constructed, representative host galaxy sample. We find that the ZTF sample of TDE hosts is dominated by compact ``green valley'' galaxies. After we restrict the comparison sample to galaxies with a similar concentration, as measured by Sersic index, we find this green valley over representation is even larger. That is, concentrated red sequence galaxies are not producing TDEs at elevated levels. We present host galaxy spectra which show that E+A galaxies are overrepresented in the ZTF sample by a factor of $\approx$22, which is lower than previous TDE host galaxy studies have found. We find that this overrepresentation can be fully accounted for when taking into account the masses, colors, and S\'ersic indices of the ZTF TDE hosts. The combination of both green colors and high S\'ersic index of the typical TDE host galaxy could be explained if the TDE rate is temporarily enhanced following a merger that leads to a higher central concentration of stars.
\end{abstract}


\section{Introduction} \label{sec:intro}
Supermassive black holes (SMBHs) are thought to reside at the center of every large galaxy, greatly influencing their environments, both in their galactic nuclei and on larger scales. However, unless a SMBH is close enough that we can precisely measure its gravitational pull on the stars and gas in its potential well or bright enough due to gas-fueled accretion, these objects are difficult to study. In distant galaxies with quiescent SMBHs, observations of tidal disruption events (TDEs) are one way to ascertain the presence and perhaps the properties of the central SMBH. 

TDEs occur when a star is kicked into an orbit that brings it close enough to the SMBH to be tidally disrupted and accreted \citep[e.g.,][]{Hills75, Frank76, Rees88}. These events are observed as bright, nuclear flares and have been discovered via observations from X-ray to optical wavelengths \citep{Saxton20, vanVelzen20b}. In order for these TDEs to be observable, the star's tidal disruption radius must be outside of the SMBH event horizon. The event horizon radius scales with black hole mass and thus, for a sun-like star, non-spinning SMBHs larger than $10^8 M_\odot$ will be too massive to host observable TDEs \citep{Hills75}. This allows for the unique opportunity to find and study low-mass SMBHs, potentially including intermediate-mass black holes, and their host galaxies, as well as accretion physics and relativistic jet formation.

There is still much debate about what types of nuclear environments are most likely to host a TDE, particularly in the mechanisms that create the fatal stellar orbits that drive these events. The environments that are likely to produce TDEs can be linked to properties that reach far beyond the nucleus of a galaxy though. \citet{Graur17} tested whether the TDE rate depends on kpc-scale global galaxy properties, including the stellar surface mass density and and the velocity dispersion, which are more easily observable than nuclear properties. They found that TDE hosts have, on average, a higher stellar mass surface density and marginally lower velocity dispersions than a control sample of galaxies. Multiple studies have shown that TDEs appear to be observed preferentially in post-starburst galaxies (otherwise known as K+A or E+A galaxies) \citep{Arcavi14, French16, LawSmith17} whose spectra have deep Balmer absorption lines but no significant [O II] emission, indicating a burst of star-formation that occurred approximately a Gyr ago that has since subsided. Several mechanisms connecting the large scale properties of the host galaxy, in this the case star formation history and global stellar population, and the dynamics of the nuclear region have been proposed. In particular, E+A galaxies are known to have high S\'ersic indices, large bulge-to-total light ratios, and high concentration indices \citep{Yang08}. The nuclear stellar overdensities caused by merger-triggered bursts of star formation in these galaxies have been shown to greatly enhance the TDE rate, possibly because these overdensities lead to a greater number of stars able to fill the loss cone of stars that can be tidally disrupted \citep{Stone16b, Stone16a, French20}. 

Not all TDEs occur in post-starburst galaxies, however. \citet{LawSmith17} studied a sample of TDE host galaxies within the context of the local galaxy population, and while the sample they used was dominated by post-starburst galaxies (3/5 used in their analysis could be classified as quiescent, Balmer-strong), they found that the majority of the TDE hosts reside in the green valley, between star-forming and passive galaxies, have bluer bulge colors, higher S\'ersic indices, and higher bulge-total-light ratios with respect to galaxies of similar masses, regardless of E+A classification. \citet{French20} studied four TDE host galaxies with high spatial resolution HST imaging: one post-starburst, two quiescent Balmer-strong galaxies, and one early type galaxy, classified by their spectra. They found that, compared to early type galaxies of similar stellar mass, the TDE host galaxies have higher central surface brightnesses and stellar mass surface densities on 30-100 pc scales, regardless of host galaxy type. Understanding the properties of not only E+A galaxies, but of the variety of galaxy types that produce TDEs is important for understanding the specific mechanisms that trigger TDEs both on large, galactic scales as well as in the nuclear environments surrounding the SMBH. In this paper, we investigate the properties of the latest Zwicky Transient Facility \citep[ZTF;][]{Bellm19, Graham19, Masci19} sample of TDE host galaxies and compare them with the properties of galaxies grouped by a variety of schema.

Previous studies that have aimed to understand TDE host galaxies have had to assemble samples from multiple surveys. This study is the first to use a systematically discovered sample of TDEs from a single survey, making the following analysis free from heterogeneous selection effects from multiple surveys. The ZTF sample selection criteria are also totally agnostic to host galaxy type, apart from rejecting galaxies that can be classified as broad-line AGN, and is therefore a prime sample for understanding properties of TDE host galaxies. See \citet{vanVelzen20} for a more detailed overview of the ZTF alert filtering and photometric selection criteria used to discover new TDEs. We then perform further follow-up of TDE candidates with spectroscopy to confirm the TDE classification and determine the TDE spectral class, discussed further in Section \ref{sec:ZTFhosts}, as well as perform UV monitoring with \textit{Swift}.

In this paper, we focus on the sample of 16 TDEs first presented in \citet{vanVelzen20} plus 3 new TDEs detected in ZTF thereafter. We study the properties of the galaxies hosting these ZTF TDEs using both photometry and spectroscopy in order to better understand the environments and mechanisms that produce them. We also compare the photometric and spectroscopic properties of these hosts to a sample of galaxies from the Sloan Digital Sky Survey (SDSS) in order to study them within the context of the local galaxy population. In Section \ref{sec:obs} we present the TDE host galaxy sample with corresponding photometric and spectroscopic data, and the SDSS comparison sample used in the following analysis. In Section \ref{sec:analysis} we present the results and methods used to obtain them. We end with a discussion presented in Section \ref{sec:discussion} and conclusions and future work in Section \ref{sec:conclusion}.

\section{Sample \& Data} \label{sec:obs}

\subsection{ZTF TDE Host Galaxies} \label{sec:ZTFhosts}
ZTF has detected 19 spectroscopically confirmed TDEs, 16 of which were originally presented in \citet{vanVelzen20}, and 3 of which that have been detected by ZTF but have not yet been reported in the literature. The light curves and spectra of these 3 new TDEs will be presented in Hammerstein et al.~(2020, in prep). The discovery history for the first 16 can be found in \citet{vanVelzen20}. We present this sample in Table \ref{tab:sample} with the redshift, host galaxy stellar mass, and TDE class. We give the IAU and ZTF names, as well as the internal name assigned to each TDE\footnote{For ease of communication, we assigned each TDE a name of a character from the popular HBO television series \textit{Game of Thrones} (GOT).}. The ZTF TDE host galaxies have redshifts in the range $0.02 \lesssim z \lesssim 0.2$, which are obtained from the spectrum of the TDE as the majority of TDE hosts do not have a pre-flare spectrum.

The TDE hosts have total stellar masses in the range $9.31 \leq \log(M_{\star}/M_\odot) \leq 10.63$. Stellar population synthesis of the pre-flare photometry, obtained from SDSS, Pan-STARRS1 (PS1), and GALEX was used to estimate the total stellar mass of each galaxy as well as obtain extinction-corrected, synthetic rest-frame colors \citep[see][]{vanVelzen20}. We adopt the same model choices as \citet{Mendel14}, a catalog we use for our comparison sample.

We have also listed the spectral class of the TDE in Table \ref{tab:sample}. The three different spectral classes correspond to emission features seen in the TDE spectrum. These classes are defined by \citet{vanVelzen20} as:
\begin{itemize}
    \item[\it i.]  {\it TDE-H}: broad H$\alpha$ and H$\beta$ emission lines. 
    
    \item[\it ii.] {\it TDE-H+He}: broad H$\alpha$ and H$\beta$ emission  lines and a broad complex of emission lines around He II $\lambda$4686. The majority of the sources in this class also show N III $\lambda$4640 and emission at $\lambda$4100 (identified as N III $\lambda$4100  instead  of  H$\delta$), plus and in some cases also O III $\lambda$3760.
    
    \item[\it iii.] {\it TDE-He}: no broad Balmer emission lines, a broad emission line near He~II~$\lambda4686$ only.
\end{itemize}

To match the procedure used for the SDSS comparison sample, we use the SDSS and PS1 calibrated, sky-subtracted, corrected $g$- and $r$-band frames for photometric measurements performed in Section \ref{sec:analysis}. The PSF at any pixel in a given SDSS frame is easily reconstructed using the corresponding psField file for that observation and the standalone code \texttt{readAtlasImages}\footnote{\href{https://www.sdss.org/dr16/software/}{https://www.sdss.org/dr16/software/}}. For PS1 images, we use \texttt{PSFEx}\footnote{\href{http://www.astromatic.net/software/psfex}{http://www.astromatic.net/software/psfex}} \citep{Bertin11} to model the PSF in each frame.

\begin{figure*}
    \centering
    \includegraphics[width=0.9\textwidth]{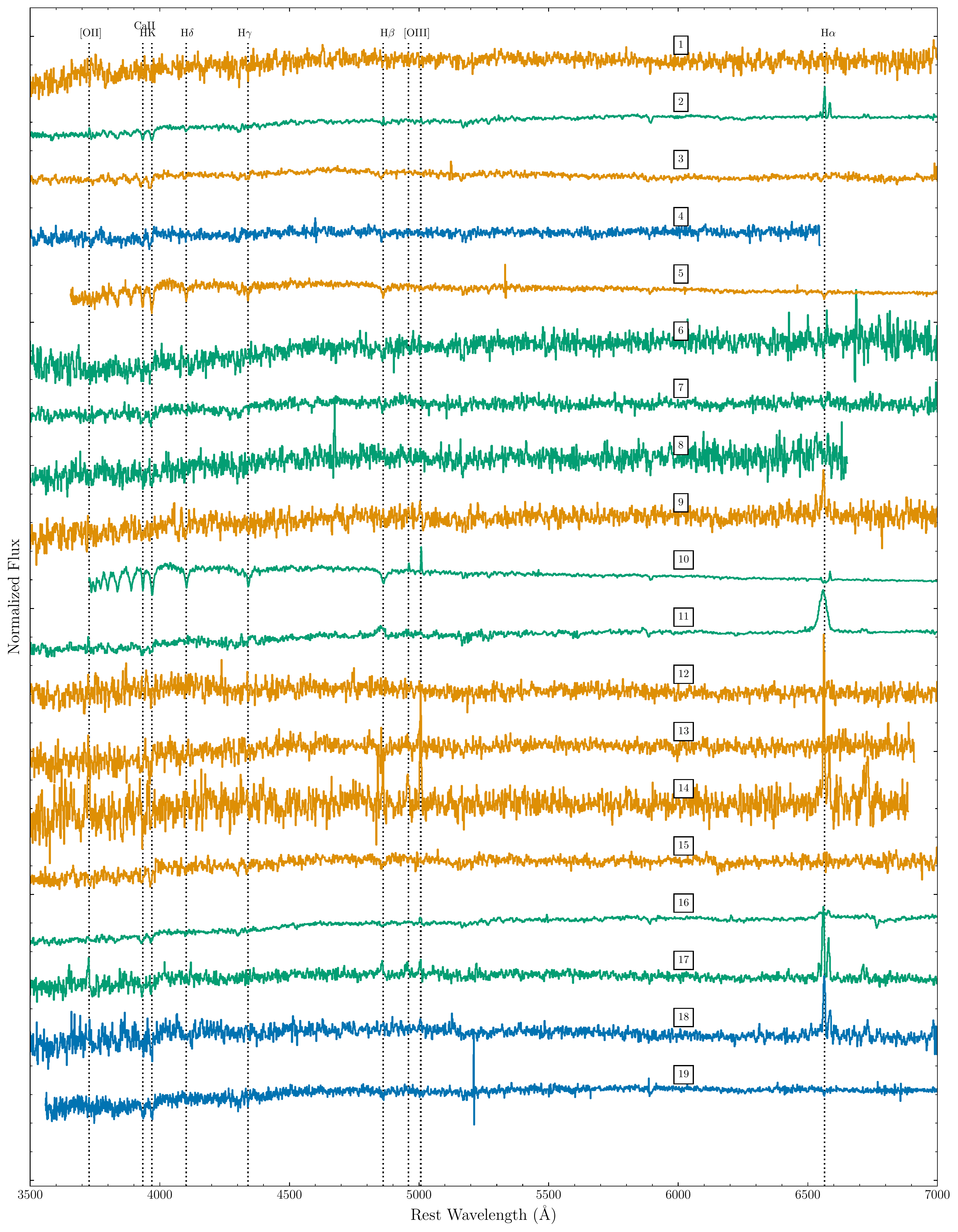}
    \caption{Rest-frame host galaxy spectra of the ZTF TDE sample. These spectra are not corrected for Galactic or internal extinction. The color of the spectrum corresponds to the TDE class from \citet{vanVelzen20}, with orange being TDE-H, blue is TDE-He, and green being TDE-H+He.}
    \label{fig:spectra}
\end{figure*}

Spectra of the host galaxies are primarily used to measure the H$\alpha$ flux and equivalent width and the Lick H$\delta_{\rm A}$ absorption index. In Table \ref{tab:sample}, we give the telescope/instrument used to obtain the spectrum. We used \texttt{PyRAF} to reduce the spectra with standard long-slit spectroscopy data reduction procedures. These spectra are presented in Figure \ref{fig:spectra}.

\begin{deluxetable*}{l l l l l r l l r r}
\tablecaption{ZTF TDE Host Galaxies}
\tablehead{ \colhead{ID} & \colhead{IAU Name} & \colhead{ZTF Name} &
\colhead{GOT Name} & \colhead{Redshift} & \colhead{$\log(M_{\star}/M_\odot$)} & \colhead{TDE Class} & \colhead{Telescope/Inst.} & \colhead{H$\alpha_{\rm em}$ EW} & \colhead{H$\delta_{\rm A}$ EW}}
\startdata
1 & AT2018zr & ZTF18aabtxvd & Ned & 0.071 & 10.02$_{0.18}^{0.09}$ & TDE-H & LDT/DeVeny & $-0.15\pm0.64$ & 3.36\\
2 & AT2018bsi & ZTF18aahqkbt & Jon & 0.051 & 10.60$_{0.06}^{0.05}$ & TDE-H+He & SDSS/BOSS & $6.08\pm0.06$ & 1.53\\
3 & AT2018hco & ZTF18abxftqm & Sansa & 0.088 & 9.93$_{0.18}^{0.09}$ & TDE-H & LDT/DeVeny & $0.70\pm0.34$ & 2.00\\
4 & AT2018iih & ZTF18acaqdaa & Jorah & 0.212 & 10.78$_{0.15}^{0.09}$ & TDE-He & LDT/DeVeny & $1.89\pm1.65$ & $-0.28$\\
5 & AT2018hyz & ZTF18acpdvos & Gendry & 0.0458 & 9.77$_{0.26}^{0.12}$ & TDE-H & SDSS/BOSS & $-0.29\pm0.14$ & 5.13\\
6 & AT2018lni & ZTF18actaqdw & Arya & 0.138 & 9.96$_{0.15}^{0.10}$ & TDE-H+He & LDT/DeVeny & $-1.82\pm0.65$ & 0.17\\
7 & AT2018lna & ZTF19aabbnzo & Cersei & 0.091 & 9.49$_{0.09}^{0.12}$ & TDE-H+He & LDT/DeVeny & $-0.36\pm0.49$& 1.84\\
8 & AT2019cho & ZTF19aakiwze & Petyr & 0.193 & 10.14$_{0.16}^{0.17}$ & TDE-H+He & LDT/DeVeny & $0.98\pm1.86$ & 1.84\\
9 & AT2019bhf & ZTF19aakswrb & Varys & 0.1206 & 10.23$_{0.12}^{0.15}$ & TDE-H & LDT/DeVeny & $12.81\pm1.35$ & 5.63\\
10 & AT2019azh & ZTF17aaazdba & Jaime & 0.022 & 9.84$_{0.14}^{0.15}$ & TDE-H+He & SDSS/BOSS & $0.77\pm0.08$ & 7.65\\
11 & AT2019dsg & ZTF19aapreis & Bran & 0.0512 & 10.37$_{0.12}^{0.17}$ & TDE-H+He & Lick/Kast & $30.63\pm0.46$ & 1.28\\
12 & AT2019ehz & ZTF19aarioci & Brienne & 0.074 & 9.69$_{0.09}^{0.15}$ & TDE-H & LDT/DeVeny & $0.40\pm1.06$ & 3.58\\
13 & AT2019mha & ZTF19abhejal & Bronn & 0.148 & 10.05$_{0.11}^{0.15}$ & TDE-H & LDT/DeVeny & $-0.55\pm0.87$ & 3.66\\
14 & AT2019meg & ZTF19abhhjcc & Margaery & 0.152 & 9.66$_{0.05}^{0.05}$ & TDE-H & LDT/DeVeny & $23.69\pm1.29$ & 2.01\\
15 & AT2019lwu & ZTF19abidbya & Robb & 0.117 & 9.86$_{0.12}^{0.15}$ & TDE-H & LDT/DeVeny & $0.27\pm0.63$ & 3.73\\
16 & AT2019qiz & ZTF19abzrhgq & Melisandre & 0.0151 & 10.04$_{0.10}^{0.14}$ & TDE-H+He & LDT/DeVeny & $2.62\pm0.12$ & 0.57\\
17 & AT2020pj\tablenotemark{a} & ZTF20aabqihu & Gilly & 0.068 & 9.99$_{0.09}^{0.17}$ & TDE-H+He & LDT/DeVeny & $18.45\pm0.68$ & 0.76\\
18 & AT2019teq\tablenotemark{b} & ZTF19accmaxo & Missandei & 0.0874 & 9.91$_{0.07}^{0.06}$ & TDE-He & LDT/DeVeny & $17.28\pm0.87$ & 2.44\\
19 & AT2020ocn\tablenotemark{c} & ZTF18aakelin & Podrick & 0.0705 & 10.10$_{0.16}^{0.17}$ & TDE-He & SDSS/BOSS & $-0.68\pm0.14$ & 0.98
\enddata
\label{tab:sample}
\tablecomments{The names, redshifts, stellar masses, TDE spectroscopic classes, and spectroscopic measurements of the ZTF sample of TDE host galaxies. Redshifts are measured from the spectrum of the TDE, as typically no pre-flare spectroscopy is available. We note that for H$\alpha$ equivalent width (EW), a positive value indicates emission. Spectra are from the 4.3m Lowell Discovery Telescope DeVeny Spectrograph (LDT/DeVeny, PI: Gezari), the 3m Lick Kast Double Spectrograph (Lick/Kast, PI: Foley), and the SDSS BOSS spectrograph.}
\tablenotetext{a}{TNS Classification Report \#7481}
\tablenotetext{b}{TNS Classification Report \#7482}
\tablenotetext{c}{ATel \#13859}
\end{deluxetable*}

\subsection{SDSS Comparison Sample}
Throughout this paper, we compare the ZTF TDE host galaxy sample to a sample of SDSS galaxies to put the ZTF sample in the context of the local galaxy population. This comparison sample is based on the \citet{Mendel14} value added catalog of bulge, disk, and total stellar mass estimates, which contains spectroscopically classified galaxies from the main SDSS galaxy sample \citep{Strauss02}. Other values are taken from the \citet{Simard11} value added catalog of bulge+disk decompositions as well as the MPA+JHU catalogs \citep{Brinchmann04}. We remove galaxies with negative flux or continuum measurements and require a median signal-to-noise ratio per pixel of greater than 10 in the MPA+JHU catalog.

In order to correct for the flux-limited nature of the SDSS spectroscopic galaxy sample and to construct a sample representative of galaxies our TDE search is sensitive to, we further limit the comparison sample by redshift. We give AT2018hco (ID 3) as an example. The absolute magnitude at peak of the AT2018hco flare light curve is $\approx$-20.1. If the ZTF detection limit is $m = 20$, ZTF can detect this flare out to a redshift of $z \approx 0.21$. If the ZTF reference image host galaxy detection limit is $m = 22$, ZTF is complete to $M \approx -18.1$ for this particular TDE. Finally, taking the SDSS spectroscopic magnitude limit to be $m = 18$, we create a comparison galaxy sub-sample with $z \leq 0.037$, which corresponds to galaxies with $M \approx -18.1$ and $m = 18$. We repeat this process for each TDE, ensuring that each TDE sub-sample has 1,000 galaxies by randomly sampling galaxies within the appropriate redshift range. Only one TDE requires oversampling of the galaxy catalog, as the redshift cut leaves less 1,000 galaxies. After applying all cuts to the comparison catalog, we are left with a sample of 19,000 galaxies. Each galaxy is weighted by 1/19 in the figures throughout this paper.

\section{Analysis \& Results} \label{sec:analysis}

\subsection{Photometry}
\begin{figure}
    \centering
    \includegraphics[width=\columnwidth]{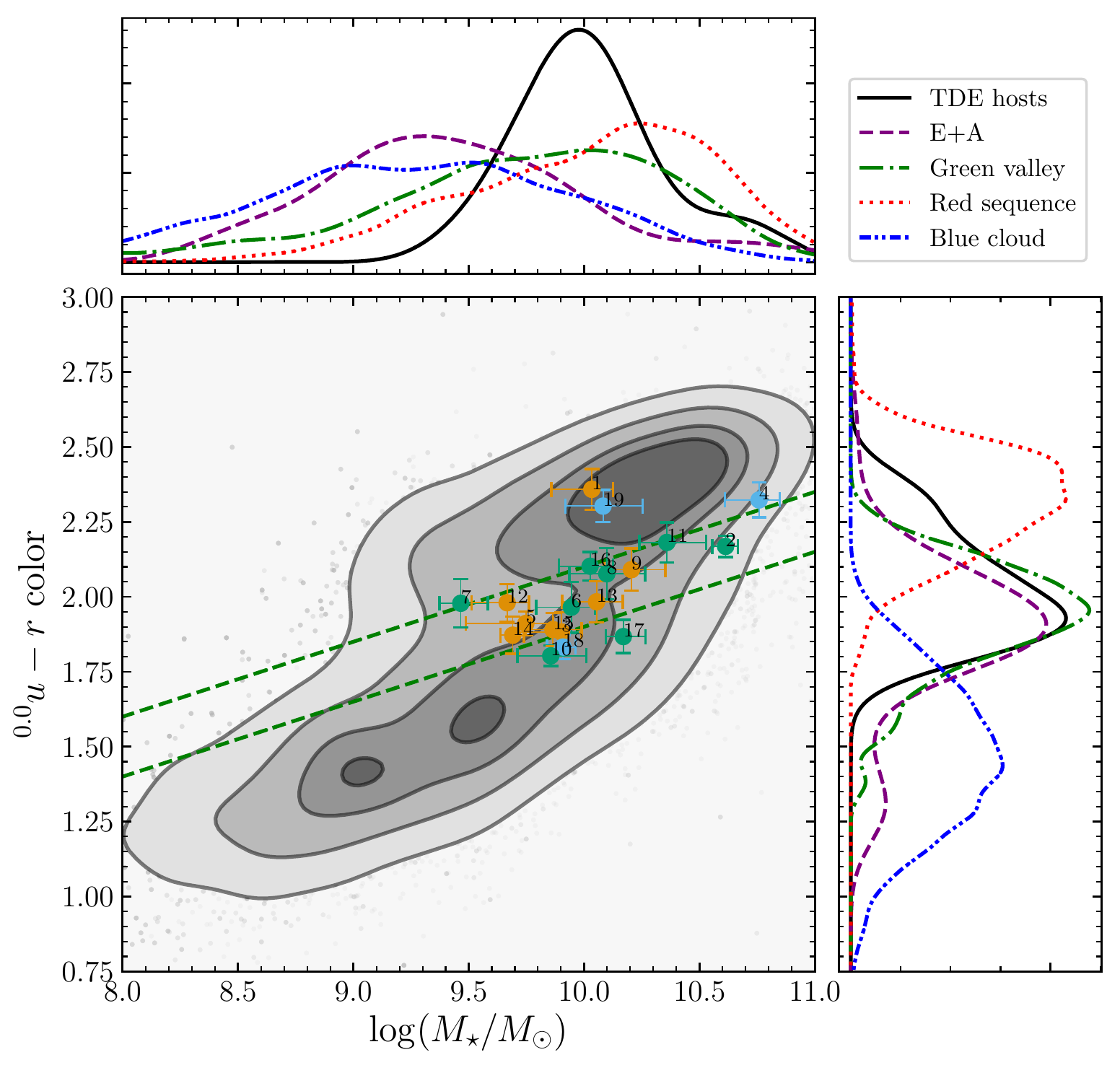}
    \caption{The extinction-corrected, synthetic rest-frame $u-r$ color of the TDE host galaxies. The green valley is denoted by the dashed green lines. TDE host galaxies are colored by their spectral type from \citet{vanVelzen20}, where orange is TDE-H, green is TDE-H+He, and blue is TDE-He. They are also numbered by the ID column in Table \ref{tab:sample}. The contours enclose a volume-limited comparison sample of galaxies, matched to the depth of ZTF, from 0.5$\sigma$ to 2$\sigma$ in steps of 0.5$\sigma$. We also show the smoothed histograms for the stellar mass and the $u-r$ color for the TDE hosts as well as the green valley, blue cloud, and red sequence. We see that the TDE hosts are generally more massive than E+A galaxies, but with similar colors characteristic of the green valley.}
    \label{fig:urcolor}
\end{figure}

We use the synthetic rest-frame and Galactic extinction corrected $u-r$ color from the stellar population synthesis of the pre-flare photometry \citep[originally presented in][]{vanVelzen20} to study where the ZTF TDE hosts fall within the red sequence, green valley, and blue cloud. Figure \ref{fig:urcolor} shows this color vs.~the total stellar mass of the TDE host galaxies against the sample of SDSS galaxies. The green valley is included on this figure, originally defined by \citet{Schawinski14}, but we redefine the upper limit here as our comparison sample has a different redshift cut:
\begin{equation}
    ^{0.0}u-r(M_{\rm gal}) = -0.40 + 0.26 \times M_{\rm gal}.
\end{equation}
We have kept a similar width to the original \citet{Schawinski14} green valley definition for the rest-frame $u-r$ color without internal dust corrections. The ZTF host galaxy sample is clearly dominated by green valley galaxies, with 63\% of the TDE hosts falling within the limits of the green valley region compared to $\sim$13\% of the SDSS comparison sample. We also include smoothed, normalized histograms of the galaxy stellar mass and $u-r$ color, for several groups of galaxies. The smoothed histograms show that the TDE hosts are typically more massive than E+A galaxies, but with similar colors characteristic of the green valley.

\begin{figure}
    \centering
    \includegraphics[width=\columnwidth]{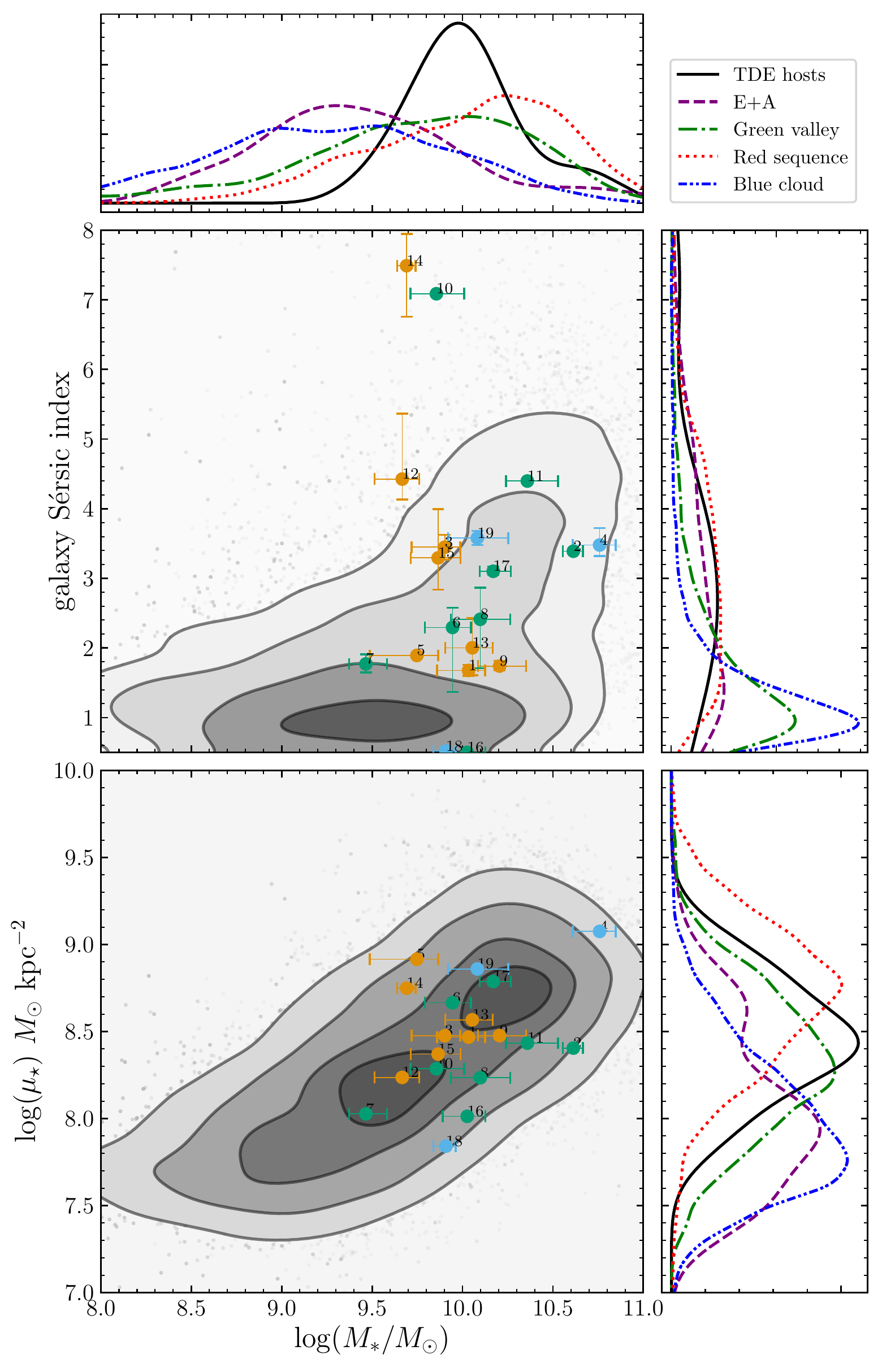}
    \caption{\textit{Top panel}: The total galaxy S\'ersic index vs.~the total stellar mass. AT2019azh (ID 10) is an extreme post-starburst galaxy and high S\'ersic index is expected. AT2019meg (ID 14) has a small angular size which likely affects the fit, resulting in the larger uncertainties and high S\'ersic index. \textit{Bottom panel}: The effective stellar mass surface density vs.~the total stellar mass. Colors of points, labels, and contours are the same as previous figures. We also show the smoothed histograms for the stellar mass, the S\'ersic index, and the effective stellar surface mass density for the TDE hosts as well as the green valley, blue cloud, and red sequence. We see that the TDE hosts have light profiles characteristic of red sequence galaxies.}
    \label{fig:sersic}
\end{figure}

To study the light profile of each host galaxy, we perform two-dimensional S\'ersic profile fits to the photometry using \texttt{GIM2D} \citep{Simard02}. Following the procedure of \citet{Simard11}, we performed simultaneous $g$- and $r$-band fits on the calibrated, sky-subtracted corrected frames of each host galaxy to obtain the total galaxy S\'ersic index. The top panel of Figure \ref{fig:sersic} shows the results of fitting the TDE hosts with a pure S\'ersic model, along with the SDSS comparison sample. Many of ZTF TDE host galaxies have profiles between a de Vaucouleurs profile ($n_g=4$) and a exponential disk profile ($n_g=1$). We also show the smoothed histograms for the stellar mass and the S\'ersic index for the TDE hosts as well as the green valley, blue cloud, red sequence, and E+A galaxies in the comparison sample. TDE hosts and E+A galaxies have steeper S\'ersic indices than the green valley, more characteristic of the red sequence.

We calculate the effective stellar surface mass density, $\mu_{\star}$,  of the TDE hosts and the comparison sample using the estimated half light radius, $\theta$, from \texttt{GIM2D}. $\mu_{\star}$ is calculated using the following equation:
\begin{equation}
    \mu_\star = M_\star / (2\pi\theta^2)~[M_\odot \mathrm{kpc}^{-2}].
\end{equation}
The bottom panel of Figure \ref{fig:sersic} shows the effective stellar surface mass density vs.~total stellar mass for the TDE hosts as well as the comparison sample. The TDE hosts have surface mass densities similar to other galaxies in the green valley.

\subsection{Spectral Measurements}
The H$\alpha$ emission equivalent width (EW) and the Lick H$\delta_{\rm A}$ absorption index can be used to explore the star formation history of a galaxy. TDE hosts in previous studies appear to be overrepresented in E+A or post-starburst galaxies, which occupy a specific region in the H$\alpha$ EW vs. H$\delta_{\rm A}$ absorption index parameter space \citep{French16, LawSmith17}. \citet{French16} isolate E+A galaxies by requiring H$\delta_{\rm A}-\sigma(\rm{H}\delta_{\rm A}) > 4.0$ and H$\alpha$ EW $<$ 3.0. These restrictions select galaxies that do not have any active star-formation, as seen from weak H$\alpha$ emission, but strong H$\delta_{\rm A}$ absorption from A stars indicates star-formation in the past $\sim$ Gyr. Both \citet{French16} and \citet{LawSmith17} also employ a looser cut, H$\delta_{\rm A} > 1.31$, allowing for host galaxies that have several possible star-formation histories. Here we make the distinction that E+A/post-starburst galaxies are identified with the stricter cut on both H$\alpha$ and H$\delta$ (hereafter E+A), while quiescent Balmer-strong galaxies are identified with the looser cut on these values (hereafter QBS). Both of these cuts are included in the following analysis.

\begin{figure}
    \centering
    \includegraphics[width=\columnwidth]{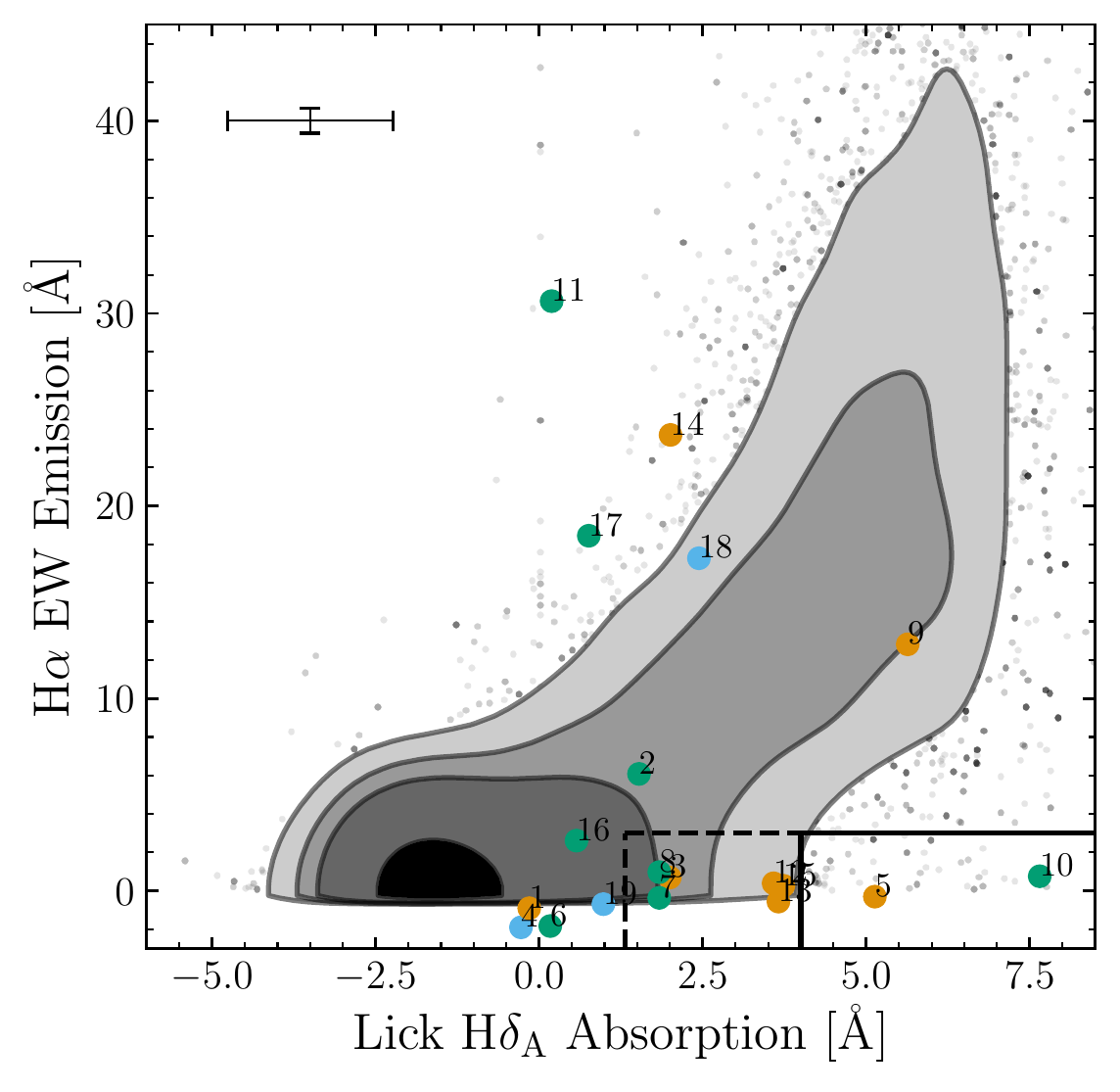}
    \caption{The H$\delta_{\rm A}$ absorption index vs.~the H$\alpha$ emission equivalent width. The median uncertainties for the TDE hosts are shown in the top left. The E+A region is the solid line while the QBS region is the dashed line. AT2019dsg (ID 11) is a star-forming galaxy. Two host galaxies fall within the strict E+A category: AT2018hyz (ID 5) and AT2019azh (ID 10). Colors of points, labels, and contours are the same as previous figures.}
    \label{fig:hdha}
\end{figure}

We fit the spectra with stellar population models using \texttt{ppxf} \citep{Cappellari17} to fit the stellar continuum and emission lines, including Balmer lines, [OII], [SII], [OIII], [OI], and [NII]. We use models from the MILES library of stellar spectra \citep{Vazdekis15}, with the models based on a standard Salpeter IMF and \citet{Girardi00} isochrones and covering the rest-frame range 3540--7410 \AA. We measure the H$\delta_{\rm A}$ absorption index using the \texttt{ppxf} best-fit stellar continuum following the procedure and bandpasses given in \citet{Worthey97}. The total H$\alpha$ EW in Angstroms is measured from the host spectrum and is corrected for Balmer absorption by subtracting the equivalent width of the absorption line in the best-fit stellar continuum from the total EW of the line, leaving only the H$\alpha$ emission EW. The H$\alpha$ line for AT2018iih (ID 4) is redshifted out of the LDT bandpass. Thus, we measure the H$\beta$ line, and take H$\alpha$ to be $\approx$3 times H$\beta$, folding this into the uncertainties as well. Since the Balmer decrement (H$\alpha$/H$\beta$) will only increase in the presence of dust extinction, assuming a ratio of $\approx$3 is a conservative estimate. Figure \ref{fig:hdha} shows the H$\delta_{\rm A}$ absorption index vs.~the H$\alpha$ emission EW. Two TDE hosts are within the bounds of the E+A region defined by \citet{French16}.

\begin{figure}
    \centering
    \includegraphics[width=\columnwidth]{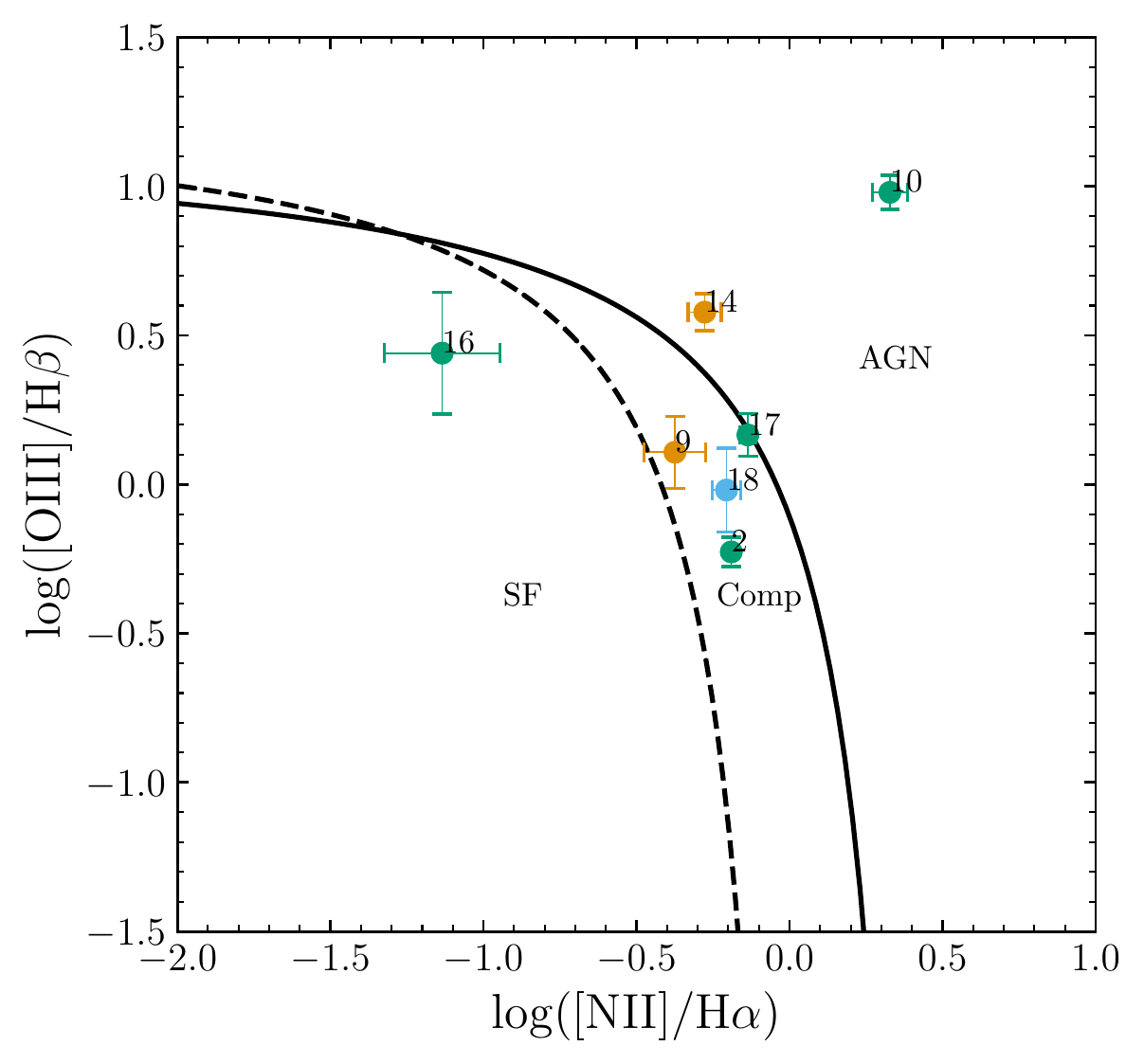}
    \caption{BPT diagram for the 7 TDE hosts that show prominent nebular emission lines. We show the separation lines of \citet{Kewley01} and \citet{Kauffmann03} as solid and dashed lines, respectively. The line ratios for the majority of the TDE hosts can be explained, at least in part, by star formation.}
    \label{fig:BPT}
\end{figure}

For the TDE hosts with prominent emission lines (7/19), we plot the emission line ratios measured with \texttt{ppxf} on a BPT diagram \citep{Baldwin81} in Figure \ref{fig:BPT}. 2 hosts (IDs 10 and 14) fall within the AGN region of the diagram, one host (ID 16) falls within the star-forming region of the diagram, while the remaining 4 (IDs 2, 9, 17, and 18) are between AGN and star-forming in the composite region. We discuss these results in Section \ref{sec:discussion}. Note that we do not include AT2019dsg despite its prominent emission lines because H$\alpha$ is rotationally broadened and we cannot resolve the [NII] doublet.

\section{Discussion} \label{sec:discussion}
Green valley galaxies dominate the ZTF TDE host galaxy sample. We find an overrepresentation of green valley galaxies of $\approx 5\times$, which is not accounted for by controlling for mass or S\'ersic index. \citet{LawSmith17} used the definition of the green valley based on total star formation rate and found that their sample of TDE host galaxies may be transitioning from star-forming to quiescent, a time during which quenching of star formation causes galaxies to cross into the green valley \citep{Schawinski14}. The green valley is also known to host a population of quiescent, Balmer-strong galaxies (including post-starburst or E+A galaxies), which previous studies observed to be overrepresented in TDE host galaxy samples \citep{Arcavi14, LawSmith17, French16}. Figure \ref{fig:hdha} reveals that two galaxies in this sample can be classified as E+A (AT2018hyz and AT2019azh, IDs 5 and 10 respectively) using the guidelines in Figure \ref{fig:hdha}. AT2019azh is an extreme post-starburst galaxy, as noted by \citet{Hinkle20}. While 2 of 19 hosts fall within the E+A region, implying an E+A fraction of 10\% with a binomial confidence interval of 1\%--33\%, this region makes up just 0.49\% of the SDSS comparison sample, implying that E+A galaxies are overrepresented in the ZTF sample by a factor of $\approx$22. The overrepresentation is lower than previous studies have found by a factor of at least 4. This overrepresentation of E+A galaxies declines to a factor of $\approx$7 when selecting only on green valley galaxies. We also make a cut on ``concentrated'' galaxies ($n_g > 2.0$), and find that the E+A overrepresentation is a factor of $\approx$15. If we combine these two criteria, then the E+A overrepresentation is $\approx 3\times$. 

To further test this result, we repeat these calculations by further limiting the comparison sample to the mass range of the TDE hosts, $9.47 \leq \log(M/M_\odot) \leq 10.76$. E+A galaxies are now overrepresented by a factor of $\approx$29. When considering only green valley galaxies in this mass limited sample, the overrepresentation declines to a factor of $\approx$8. The S\'ersic index cut gives an E+A overrepresenation of $\approx 29\times$. After combining these two criteria, we find that the E+A overrepresentation can be completely accounted for. These calculations can also be repeated to include the QBS region of the H$\alpha$-H$\delta_{\rm A}$ figure, in order to account for more ambiguous star-formation histories. These numbers are presented in Table \ref{tab:overrep}.

E+A galaxies are also known to have nuclear stellar overdensities and thus have a higher S\'ersic index \citep[e.g.~][]{French20}. \citet{LawSmith17} concluded that the S\'ersic index is a stronger factor in enhancing the TDE rate in a galaxy than other properties that they measured. Figure \ref{fig:sersic} shows that galaxies in the ZTF sample have higher S\'ersic indices than is typical for galaxies of similar stellar mass. \citet{LawSmith17} found a similar trend in S\'ersic index for TDE host galaxies with S\'ersic indices in the range $1 < n_g < 5$. We note, however, that the resolution of SDSS and PS1 is insufficient to fit the bulge S\'ersic index independently of a disk that may be present. Thus, if a substantial disk component is present in a galaxy, the bulge component may actually have a higher S\'ersic index. These fits should therefore be interpreted as lower limits on the bulge S\'ersic index.

Despite the insufficient resolution of SDSS, \citet{Simard11} found that galaxies fit with a free $n_b$ bulge+disk model resulting in a bulge-to-total light ratio of $0.2 \leq (B/T) \leq 0.45$ required a bulge+disk model while a fit resulting in $(B/T) > 0.75$ did not require a bulge+disk model to fit the light profile. We performed free $n_b$ bulge+disk decompositions on the sample of ZTF host galaxies in order to determine which galaxies may require a bulge+disk model to properly fit the light profile. Using the criteria from \citet{Simard11}, 6 of the TDE host galaxies require a bulge+disk model to properly fit the light profile and 5 require only a S\'ersic profile, while the rest remain ambiguous. The ZTF TDE hosts may have steeper bulge profiles than Figure \ref{fig:sersic} demonstrates but higher resolution imaging is needed to separate the bulge and disk components.

\citet{French17} constructed a BPT diagram for 5 TDE hosts, finding that the location of the TDE hosts in AGN region of the diagram is consistent with a sample of quiescent, Balmer-strong galaxies from SDSS. Figure \ref{fig:BPT} shows a BPT diagram for the TDE hosts with prominent nebular emission lines. One of the hosts (ID 16) has emission line ratios consistent with only star-formation. While previous studies, such as \citet{French17}, found that several TDE hosts had emission line ratios consistent with AGN or LINER-like activity, only 2 of the TDE hosts (IDs 10 and 16) have emission line ratios consistent with the AGN region of the BPT diagram. The remaining 4 hosts have emission line ratios that may, in part, be attributed to star formation.  For TDE hosts falling in the AGN region of the BPT diagram, \citet{French17} mention several possible ionization mechanisms which may act to enhance the TDE rate in galaxies. These mechanisms include a low-luminosity AGN fueled by a circumnuclear gas reservoir and shocks resulting from a recent merger or starburst.

Figures \ref{fig:urcolor} and \ref{fig:sersic} show smoothed histograms for the $u-r$, galaxy stellar mass, and S\'ersic index. From Figure \ref{fig:urcolor}, we see that TDE hosts are typically more massive than E+A galaxies, which are on the low end of the green valley in terms of mass. We also see that TDE host galaxies are in the green valley in terms of color, with a distribution that matches that of E+A galaxies. Figure \ref{fig:sersic} shows that both E+A galaxies and TDE host galaxies have S\'ersic indices different from other galaxies in the green valley, with a distributions closer to that of the red sequence.

\citet{Schawinski10} found that the migration of low-mass, early-type galaxies from the blue cloud to the green valley is linked to mergers but the ability to link merger signatures to these galaxies over timescales longer than $\sim$500 Myr post-merger is limited. Subsequently, \citet{Schawinski14} found that there are two main causes for galaxies to fall in the green valley. They found that morphologically late-type galaxies in the green valley are consistent with a scenario where the supply of gas fueling star-formation is shut off, leading to an exhaustion of the remaining gas over the next Gyr or so. Morphologically early-type galaxies are in the green valley as a result of a scenario where quenching of star formation happened rapidly and was accompanied by a change in morphology from disk to spheroid, likely as a consequence of a merger. E+A galaxies are thought to have undergone merger-triggered bursts of star-formation that place them in the green valley and that lead to centrally concentrated stellar distributions \citep{Yang08}. Indeed, \citet{Schawinski14} found that morphologically early-type galaxies in the green valley show classic post-starburst stellar populations.

Given that the TDE hosts show a distribution in S\'ersic index more similar to red, early-type galaxies, it is possible that TDE host galaxies, E+A or not, are more likely to come from galaxies that have undergone some type of merger that produces centrally concentrated stellar distributions and which places them in the green valley. \citet{Prieto16} found that the host galaxy of TDE ASASSN-14li \citep{Holoien16}, PGC 043234, possesses properties indicative of a recent merger, including AGN activity, post-starburst populations, and emission line filaments extending up to 10 kpc from the galaxy itself. This further supports the idea that E+A galaxies are the result of galaxy mergers that could enhance the TDE rate. The E+A phase of a post-merger system could also be a time when the TDE rate is greatly enhanced as compared to other phases of a post-merger system, which would explain the overrepresentation. \citet{Stone18} showed the nuclear stellar overdensities created as the result of starbursts can reasonably match the TDE delay-time distribution and that the post-starburst TDE rate does indeed decline with time. This may be a reason why we find few red galaxies in the ZTF TDE sample. The majority of the ZTF TDE hosts are not E+A galaxies, but they do have more centrally concentrated stellar distributions. We propose that, similar to \citet{LawSmith17}, the overall stellar distribution in a galaxy is more important than the E+A classification and that E+A galaxies are only a subset of the larger population of galaxies that are likely to host TDEs.

\begin{deluxetable*}{l c c c c}
\tablecaption{E+A \& QBS Overrepresentation}
\tablehead{ & \colhead{Overall} & \thead{Green \\ Valley} &  \colhead{$n_g > 2.0$} & \thead{Green Valley \\ + $n_g > 2.0$}}
\tablecolumns{5}
\startdata
Full Sample (E+A) & 22$\times$ & 7$\times$ & 15$\times$ & 3$\times$ \\
$9.47 \leq \log(M_\star/M_\odot) \leq 10.76$ (E+A) & 29$\times$ & 8$\times$ & 29$\times$ & 1$\times$ \\
Full Sample (QBS) & 16$\times$ & 10$\times$ & 13$\times$ & 6$\times$\\
$9.47 \leq \log(M_\star/M_\odot) \leq 10.76$ (QBS) & 17$\times$ & 9$\times$ & 21$\times$ & 3$\times$ 
\enddata
\label{tab:overrep}
\tablecomments{The E+A overrepresentation in the ZTF TDE host galaxy sample calculated with respect to the full galaxy comparison sample and a mass-limited comparison sample, for the E+A criteria and the QBS criteria. We give the overall overrepresentation, the overrepresentation when considering only the green valley as well as considering only concentrated galaxies, and the overrepresentation when considering galaxies with S\'ersic index ($n_g$) greater than or equal to the median S\'ersic index of the TDE host galaxies.}
\end{deluxetable*}

\section{Conclusions} \label{sec:conclusion}
We have studied a sample of galaxies hosting TDEs detected by ZTF in the first two-thirds of survey operations. Our main conclusions are:
\begin{itemize}
    \item The ZTF TDE host galaxy sample is dominated by green valley galaxies, with 63\% of the TDE hosts having $u-r$ colors corresponding to the green valley, compared to only 13\% of the comparison sample of galaxies.
    \item E+A galaxies, which we define spectroscopically, are overrepresented in the ZTF TDE host galaxy sample by a factor of $\approx$22 compared to the SDSS comparison sample of galaxies. This overrepresentation reduces to a factor of $\approx$7 when selecting only on green valley galaxies (defined photometrically) and to a factor of $\approx$3 when selecting on green valley galaxies with concentrated stellar distributions. The apparent E+A preference for TDE host galaxies is completely accounted for when looking at galaxy populations with similar masses, colors, and S\'ersic indices as the ZTF TDE hosts.
    \item The ZTF TDE hosts have higher S\'ersic indices than galaxies of similar stellar masses and show a distribution of S\'ersic indices similar to E+A galaxies and red sequence galaxies, rather than green valley galaxies.
    \item TDE hosts may be more likely to be found in the subset of galaxies that have undergone a more recent merger that produced centrally-concentrated stellar distributions, enhancing the TDE rate.
\end{itemize}

\acknowledgments
These results made use of the Lowell Discovery Telescope at Lowell Observatory. Lowell is a private, nonprofit institution dedicated to astrophysical research and public appreciation of astronomy and operates the LDT in partnership with Boston University, the University of Maryland, the University of Toledo, Northern Arizona University and Yale University. The upgrade of the DeVeny optical spectrograph has been funded by a generous grant from John and Ginger Giovale.

Based on observations obtained with the Samuel Oschin Telescope 48-inch and the 60-inch Telescope at the Palomar Observatory as part of the Zwicky Transient Facility project. ZTF is supported by the National Science Foundation under Grant No. AST-1440341 and a collaboration including Caltech, IPAC, the Weizmann Institute for Science, the Oskar Klein Center at Stockholm University, the University of Maryland, the University of Washington, Deutsches Elektronen-Synchrotron and Humboldt University, Los Alamos National Laboratories, the TANGO Consortium of Taiwan, the University of Wisconsin at Milwaukee, and Lawrence Berkeley National Laboratories. Operations are conducted by COO, IPAC, and UW. This work was supported by the GROWTH project funded by the National Science Foundation under Grant No 1545949.

Funding for the SDSS and SDSS-II has been provided by the Alfred P. Sloan Foundation, the Participating Institutions, the National Science Foundation, the U.S. Department of Energy, the National Aeronautics and Space Administration, the Japanese Monbukagakusho, the Max Planck Society, and the Higher Education Funding Council for England. The SDSS Web Site is \url{http://www.sdss.org/}. The SDSS is managed by the Astrophysical Research Consortium for the Participating Institutions. The Participating Institutions are the American Museum of Natural History, Astrophysical Institute Potsdam, University of Basel, University of Cambridge, Case Western Reserve University, University of Chicago, Drexel University, Fermilab, the Institute for Advanced Study, the Japan Participation Group, Johns Hopkins University, the Joint Institute for Nuclear Astrophysics, the Kavli Institute for Particle Astrophysics and Cosmology, the Korean Scientist Group, the Chinese Academy of Sciences (LAMOST), Los Alamos National Laboratory, the Max-Planck-Institute for Astronomy (MPIA), the Max-Planck-Institute for Astrophysics (MPA), New Mexico State University, Ohio State University, University of Pittsburgh, University of Portsmouth, Princeton University, the United States Naval Observatory, and the University of Washington.

The Pan-STARRS1 Surveys (PS1) and the PS1 public science archive have been made possible through contributions by the Institute for Astronomy, the University of Hawaii, the Pan-STARRS Project Office, the Max-Planck Society and its participating institutes, the Max Planck Institute for Astronomy, Heidelberg and the Max Planck Institute for Extraterrestrial Physics, Garching, The Johns Hopkins University, Durham University, the University of Edinburgh, the Queen's University Belfast, the Harvard-Smithsonian Center for Astrophysics, the Las Cumbres Observatory Global Telescope Network Incorporated, the National Central University of Taiwan, the Space Telescope Science Institute, the National Aeronautics and Space Administration under Grant No. NNX08AR22G issued through the Planetary Science Division of the NASA Science Mission Directorate, the National Science Foundation Grant No. AST-1238877, the University of Maryland, Eotvos Lorand University (ELTE), the Los Alamos National Laboratory, and the Gordon and Betty Moore Foundation.

SG is funded in part by NSF CAREER grant 1454816, and NASA Swift grant 80NSSC20K0961. The UCSC team is supported in part by NSF grant AST-1518052; the Gordon \& Betty Moore Foundation; the Heising-Simons Foundation; and by a fellowship from the David and Lucile Packard Foundation to R.J.F.

\software{Astropy \citep{2018AJ....156..123A}, SciPy \citep{Virtanen_2020}, ppxf \citep{Cappellari17}, GIM2D \citep{Simard02}, PSFex \citep{Bertin11}, Source-Extractor \citep{Bertin96}.}

\bibstyle{aasjournals}
\bibliography{syp}

\begin{thebibliography}{}
\expandafter\ifx\csname natexlab\endcsname\relax\def\natexlab#1{#1}\fi
\providecommand{\url}[1]{\href{#1}{#1}}

\bibitem[{{Arcavi} {et~al.}(2014){Arcavi}, {Gal-Yam}, {Sullivan}, {Pan},
  {Cenko}, {Horesh}, {Ofek}, {De Cia}, {Yan}, {Yang}, {Howell}, {Tal},
  {Kulkarni}, {Tendulkar}, {Tang}, {Xu}, {Sternberg}, {Cohen}, {Bloom},
  {Nugent}, {Kasliwal}, {Perley}, {Quimby}, {Miller}, {Theissen}, \&
  {Laher}}]{Arcavi14}
{Arcavi}, I., {Gal-Yam}, A., {Sullivan}, M., {et~al.} 2014, \apj, 793, 38

\bibitem[{{Astropy Collaboration} {et~al.}(2018){Astropy Collaboration},
  {Price-Whelan}, {Sip{\H o}cz}, {G{\"u}nther}, {Lim}, {Crawford}, {Conseil},
  {Shupe}, {Craig}, {Dencheva}, {Ginsburg}, {VanderPlas}, {Bradley},
  {P{\'e}rez-Su{\'a}rez}, {de Val-Borro}, {Aldcroft}, {Cruz}, {Robitaille},
  {Tollerud}, {Ardelean}, {Babej}, {Bach}, {Bachetti}, {Bakanov}, {Bamford},
  {Barentsen}, {Barmby}, {Baumbach}, {Berry}, {Biscani}, {Boquien}, {Bostroem},
  {Bouma}, {Brammer}, {Bray}, {Breytenbach}, {Buddelmeijer}, {Burke},
  {Calderone}, {Cano Rodr{\'{\i}}guez}, {Cara}, {Cardoso}, {Cheedella},
  {Copin}, {Corrales}, {Crichton}, {D'Avella}, {Deil}, {Depagne}, {Dietrich},
  {Donath}, {Droettboom}, {Earl}, {Erben}, {Fabbro}, {Ferreira}, {Finethy},
  {Fox}, {Garrison}, {Gibbons}, {Goldstein}, {Gommers}, {Greco}, {Greenfield},
  {Groener}, {Grollier}, {Hagen}, {Hirst}, {Homeier}, {Horton}, {Hosseinzadeh},
  {Hu}, {Hunkeler}, {Ivezi{\'c}}, {Jain}, {Jenness}, {Kanarek}, {Kendrew},
  {Kern}, {Kerzendorf}, {Khvalko}, {King}, {Kirkby}, {Kulkarni}, {Kumar},
  {Lee}, {Lenz}, {Littlefair}, {Ma}, {Macleod}, {Mastropietro}, {McCully},
  {Montagnac}, {Morris}, {Mueller}, {Mumford}, {Muna}, {Murphy}, {Nelson},
  {Nguyen}, {Ninan}, {N{\"o}the}, {Ogaz}, {Oh}, {Parejko}, {Parley}, {Pascual},
  {Patil}, {Patil}, {Plunkett}, {Prochaska}, {Rastogi}, {Reddy Janga},
  {Sabater}, {Sakurikar}, {Seifert}, {Sherbert}, {Sherwood-Taylor}, {Shih},
  {Sick}, {Silbiger}, {Singanamalla}, {Singer}, {Sladen}, {Sooley},
  {Sornarajah}, {Streicher}, {Teuben}, {Thomas}, {Tremblay}, {Turner},
  {Terr{\'o}n}, {van Kerkwijk}, {de la Vega}, {Watkins}, {Weaver}, {Whitmore},
  {Woillez}, {Zabalza}, \& {Astropy Contributors}}]{2018AJ....156..123A}
{Astropy Collaboration}, {Price-Whelan}, A.~M., {Sip{\H o}cz}, B.~M., {et~al.}
  2018, \aj, 156, 123

\bibitem[{{Baldwin} {et~al.}(1981){Baldwin}, {Phillips}, \&
  {Terlevich}}]{Baldwin81}
{Baldwin}, J.~A., {Phillips}, M.~M., \& {Terlevich}, R. 1981, \pasp, 93, 5

\bibitem[{{Bellm} {et~al.}(2019){Bellm}, {Kulkarni}, {Barlow}, {Feindt},
  {Graham}, {Goobar}, {Kupfer}, {Ngeow}, {Nugent}, {Ofek}, {Prince}, {Riddle},
  {Walters}, \& {Ye}}]{Bellm19}
{Bellm}, E.~C., {Kulkarni}, S.~R., {Barlow}, T., {et~al.} 2019, \pasp, 131,
  068003

\bibitem[{{Bertin}(2011)}]{Bertin11}
{Bertin}, E. 2011, Astronomical Society of the Pacific Conference Series, Vol.
  442, {Automated Morphometry with SExtractor and PSFEx}, ed. I.~N. {Evans},
  A.~{Accomazzi}, D.~J. {Mink}, \& A.~H. {Rots}, 435

\bibitem[{{Bertin} \& {Arnouts}(1996)}]{Bertin96}
{Bertin}, E., \& {Arnouts}, S. 1996, \aaps, 117, 393

\bibitem[{{Brinchmann} {et~al.}(2004){Brinchmann}, {Charlot}, {White},
  {Tremonti}, {Kauffmann}, {Heckman}, \& {Brinkmann}}]{Brinchmann04}
{Brinchmann}, J., {Charlot}, S., {White}, S.~D.~M., {et~al.} 2004, \mnras, 351,
  1151

\bibitem[{{Cappellari}(2017)}]{Cappellari17}
{Cappellari}, M. 2017, \mnras, 466, 798

\bibitem[{{Frank} \& {Rees}(1976)}]{Frank76}
{Frank}, J., \& {Rees}, M.~J. 1976, \mnras, 176, 633

\bibitem[{{French} {et~al.}(2016){French}, {Arcavi}, \& {Zabludoff}}]{French16}
{French}, K.~D., {Arcavi}, I., \& {Zabludoff}, A. 2016, \apjl, 818, L21

\bibitem[{{French} {et~al.}(2017){French}, {Arcavi}, \& {Zabludoff}}]{French17}
---. 2017, \apj, 835, 176

\bibitem[{{French} {et~al.}(2020){French}, {Arcavi}, {Zabludoff}, {Stone},
  {Hiramatsu}, {van Velzen}, {McCully}, \& {Jiang}}]{French20}
{French}, K.~D., {Arcavi}, I., {Zabludoff}, A.~I., {et~al.} 2020, arXiv
  e-prints, arXiv:2002.02498

\bibitem[{{Girardi} {et~al.}(2000){Girardi}, {Bressan}, {Bertelli}, \&
  {Chiosi}}]{Girardi00}
{Girardi}, L., {Bressan}, A., {Bertelli}, G., \& {Chiosi}, C. 2000, \aaps, 141,
  371

\bibitem[{{Graham} {et~al.}(2019){Graham}, {Kulkarni}, {Bellm}, {Adams},
  {Barbarino}, {Blagorodnova}, {Bodewits}, {Bolin}, {Brady}, {Cenko}, {Chang},
  {Coughlin}, {De}, {Eadie}, {Farnham}, {Feindt}, {Franckowiak}, {Fremling},
  {Gezari}, {Ghosh}, {Goldstein}, {Golkhou}, {Goobar}, {Ho}, {Huppenkothen},
  {Ivezi{\'c}}, {Jones}, {Juric}, {Kaplan}, {Kasliwal}, {Kelley}, {Kupfer},
  {Lee}, {Lin}, {Lunnan}, {Mahabal}, {Miller}, {Ngeow}, {Nugent}, {Ofek},
  {Prince}, {Rauch}, {van Roestel}, {Schulze}, {Singer}, {Sollerman}, {Taddia},
  {Yan}, {Ye}, {Yu}, {Barlow}, {Bauer}, {Beck}, {Belicki}, {Biswas}, {Brinnel},
  {Brooke}, {Bue}, {Bulla}, {Burruss}, {Connolly}, {Cromer}, {Cunningham},
  {Dekany}, {Delacroix}, {Desai}, {Duev}, {Feeney}, {Flynn}, {Frederick},
  {Gal-Yam}, {Giomi}, {Groom}, {Hacopians}, {Hale}, {Helou}, {Henning},
  {Hover}, {Hillenbrand}, {Howell}, {Hung}, {Imel}, {Ip}, {Jackson}, {Kaspi},
  {Kaye}, {Kowalski}, {Kramer}, {Kuhn}, {Landry}, {Laher}, {Mao}, {Masci},
  {Monkewitz}, {Murphy}, {Nordin}, {Patterson}, {Penprase}, {Porter},
  {Rebbapragada}, {Reiley}, {Riddle}, {Rigault}, {Rodriguez}, {Rusholme}, {van
  Santen}, {Shupe}, {Smith}, {Soumagnac}, {Stein}, {Surace}, {Szkody}, {Terek},
  {Van Sistine}, {van Velzen}, {Vestrand}, {Walters}, {Ward}, {Zhang}, \&
  {Zolkower}}]{Graham19}
{Graham}, M.~J., {Kulkarni}, S.~R., {Bellm}, E.~C., {et~al.} 2019, \pasp, 131,
  078001

\bibitem[{{Graur} {et~al.}(2018){Graur}, {French}, {Zahid}, {Guillochon},
  {Mandel}, {Auchettl}, \& {Zabludoff}}]{Graur17}
{Graur}, O., {French}, K.~D., {Zahid}, H.~J., {et~al.} 2018, \apj, 853, 39

\bibitem[{{Hills}(1975)}]{Hills75}
{Hills}, J.~G. 1975, \nat, 254, 295

\bibitem[{{Hinkle} {et~al.}(2020){Hinkle}, {Holoien}, {Auchettl}, {Shappee},
  {Neustadt}, {Payne}, {Brown}, {Kochanek}, {Stanek}, {Graham}, {Tucker}, {Do},
  {Anderson}, {Bose}, {Chen}, {Coulter}, {Dimitriadis}, {Dong}, {Foley},
  {Huber}, {Hung}, {Kilpatrick}, {Pignata}, {Prieto}, {Rojas-Bravo}, {Siebert},
  {Stalder}, {Thompson}, {Tonry}, {Vallely}, \& {Wisniewski}}]{Hinkle20}
{Hinkle}, J.~T., {Holoien}, T.~W.~S., {Auchettl}, K., {et~al.} 2020, arXiv
  e-prints, arXiv:2006.06690

\bibitem[{{Holoien} {et~al.}(2016){Holoien}, {Kochanek}, {Prieto}, {Stanek},
  {Dong}, {Shappee}, {Grupe}, {Brown}, {Basu}, {Beacom}, {Bersier},
  {Brimacombe}, {Danilet}, {Falco}, {Guo}, {Jose}, {Herczeg}, {Long},
  {Pojmanski}, {Simonian}, {Szczygie{\l}}, {Thompson}, {Thorstensen}, {Wagner},
  \& {Wo{\'z}niak}}]{Holoien16}
{Holoien}, T.~W.~S., {Kochanek}, C.~S., {Prieto}, J.~L., {et~al.} 2016, \mnras,
  455, 2918

\bibitem[{{Kauffmann} {et~al.}(2003){Kauffmann}, {Heckman}, {Tremonti},
  {Brinchmann}, {Charlot}, {White}, {Ridgway}, {Brinkmann}, {Fukugita}, {Hall},
  {Ivezi{\'c}}, {Richards}, \& {Schneider}}]{Kauffmann03}
{Kauffmann}, G., {Heckman}, T.~M., {Tremonti}, C., {et~al.} 2003, \mnras, 346,
  1055

\bibitem[{{Kewley} {et~al.}(2001){Kewley}, {Dopita}, {Sutherland}, {Heisler},
  \& {Trevena}}]{Kewley01}
{Kewley}, L.~J., {Dopita}, M.~A., {Sutherland}, R.~S., {Heisler}, C.~A., \&
  {Trevena}, J. 2001, \apj, 556, 121

\bibitem[{{Law-Smith} {et~al.}(2017){Law-Smith}, {Ramirez-Ruiz}, {Ellison}, \&
  {Foley}}]{LawSmith17}
{Law-Smith}, J., {Ramirez-Ruiz}, E., {Ellison}, S.~L., \& {Foley}, R.~J. 2017,
  \apj, 850, 22

\bibitem[{{Masci} {et~al.}(2019){Masci}, {Laher}, {Rusholme}, {Shupe}, {Groom},
  {Surace}, {Jackson}, {Monkewitz}, {Beck}, {Flynn}, {Terek}, {Landry},
  {Hacopians}, {Desai}, {Howell}, {Brooke}, {Imel}, {Wachter}, {Ye}, {Lin},
  {Cenko}, {Cunningham}, {Rebbapragada}, {Bue}, {Miller}, {Mahabal}, {Bellm},
  {Patterson}, {Juri{\'c}}, {Golkhou}, {Ofek}, {Walters}, {Graham}, {Kasliwal},
  {Dekany}, {Kupfer}, {Burdge}, {Cannella}, {Barlow}, {Van Sistine}, {Giomi},
  {Fremling}, {Blagorodnova}, {Levitan}, {Riddle}, {Smith}, {Helou}, {Prince},
  \& {Kulkarni}}]{Masci19}
{Masci}, F.~J., {Laher}, R.~R., {Rusholme}, B., {et~al.} 2019, \pasp, 131,
  018003

\bibitem[{{Mendel} {et~al.}(2014){Mendel}, {Simard}, {Palmer}, {Ellison}, \&
  {Patton}}]{Mendel14}
{Mendel}, J.~T., {Simard}, L., {Palmer}, M., {Ellison}, S.~L., \& {Patton},
  D.~R. 2014, \apjs, 210, 3

\bibitem[{{Prieto} {et~al.}(2016){Prieto}, {Kr{\"u}hler}, {Anderson},
  {Galbany}, {Kochanek}, {Aquino}, {Brown}, {Dong}, {F{\"o}rster}, {Holoien},
  {Kuncarayakti}, {Maureira}, {Rosales-Ortega}, {S{\'a}nchez}, {Shappee}, \&
  {Stanek}}]{Prieto16}
{Prieto}, J.~L., {Kr{\"u}hler}, T., {Anderson}, J.~P., {et~al.} 2016, \apjl,
  830, L32

\bibitem[{{Rees}(1988)}]{Rees88}
{Rees}, M.~J. 1988, \nat, 333, 523

\bibitem[{{Saxton} {et~al.}(2020){Saxton}, {Komossa}, {Auchettl}, \&
  {Jonker}}]{Saxton20}
{Saxton}, R., {Komossa}, S., {Auchettl}, K., \& {Jonker}, P.~G. 2020, \ssr,
  216, 85

\bibitem[{{Schawinski} {et~al.}(2010){Schawinski}, {Dowlin}, {Thomas}, {Urry},
  \& {Edmondson}}]{Schawinski10}
{Schawinski}, K., {Dowlin}, N., {Thomas}, D., {Urry}, C.~M., \& {Edmondson}, E.
  2010, \apjl, 714, L108

\bibitem[{{Schawinski} {et~al.}(2014){Schawinski}, {Urry}, {Simmons},
  {Fortson}, {Kaviraj}, {Keel}, {Lintott}, {Masters}, {Nichol}, {Sarzi},
  {Skibba}, {Treister}, {Willett}, {Wong}, \& {Yi}}]{Schawinski14}
{Schawinski}, K., {Urry}, C.~M., {Simmons}, B.~D., {et~al.} 2014, \mnras, 440,
  889

\bibitem[{{Simard} {et~al.}(2011){Simard}, {Mendel}, {Patton}, {Ellison}, \&
  {McConnachie}}]{Simard11}
{Simard}, L., {Mendel}, J.~T., {Patton}, D.~R., {Ellison}, S.~L., \&
  {McConnachie}, A.~W. 2011, \apjs, 196, 11

\bibitem[{{Simard} {et~al.}(2002){Simard}, {Willmer}, {Vogt}, {Sarajedini},
  {Phillips}, {Weiner}, {Koo}, {Im}, {Illingworth}, \& {Faber}}]{Simard02}
{Simard}, L., {Willmer}, C. N.~A., {Vogt}, N.~P., {et~al.} 2002, \apjs, 142, 1

\bibitem[{{Stone} {et~al.}(2018){Stone}, {Generozov}, {Vasiliev}, \&
  {Metzger}}]{Stone18}
{Stone}, N.~C., {Generozov}, A., {Vasiliev}, E., \& {Metzger}, B.~D. 2018,
  \mnras, 480, 5060

\bibitem[{{Stone} \& {Metzger}(2016)}]{Stone16b}
{Stone}, N.~C., \& {Metzger}, B.~D. 2016, \mnras, 455, 859

\bibitem[{{Stone} \& {van Velzen}(2016)}]{Stone16a}
{Stone}, N.~C., \& {van Velzen}, S. 2016, \apjl, 825, L14

\bibitem[{{Strauss} {et~al.}(2002){Strauss}, {Weinberg}, {Lupton}, {Narayanan},
  {Annis}, {Bernardi}, {Blanton}, {Burles}, {Connolly}, {Dalcanton}, {Doi},
  {Eisenstein}, {Frieman}, {Fukugita}, {Gunn}, {Ivezi{\'c}}, {Kent}, {Kim},
  {Knapp}, {Kron}, {Munn}, {Newberg}, {Nichol}, {Okamura}, {Quinn}, {Richmond},
  {Schlegel}, {Shimasaku}, {SubbaRao}, {Szalay}, {Vanden Berk}, {Vogeley},
  {Yanny}, {Yasuda}, {York}, \& {Zehavi}}]{Strauss02}
{Strauss}, M.~A., {Weinberg}, D.~H., {Lupton}, R.~H., {et~al.} 2002, \aj, 124,
  1810

\bibitem[{{van Velzen} {et~al.}(2020{\natexlab{a}}){van Velzen}, {Holoien},
  {Onori}, {Hung}, \& {Arcavi}}]{vanVelzen20b}
{van Velzen}, S., {Holoien}, T. W.~S., {Onori}, F., {Hung}, T., \& {Arcavi}, I.
  2020{\natexlab{a}}, arXiv e-prints, arXiv:2008.05461

\bibitem[{{van Velzen} {et~al.}(2020{\natexlab{b}}){van Velzen}, {Gezari},
  {Hammerstein}, {Roth}, {Frederick}, {Ward}, {Hung}, {Cenko}, {Stein},
  {Perley}, {Taggart}, {Sollerman}, {Andreoni}, {Bellm}, {Brinnel}, {De},
  {Dekany}, {Feeney}, {Foley}, {Fremling}, {Giomi}, {Golkhou}, {Ho},
  {Kasliwal}, {Kilpatrick}, {Kulkarni}, {Kupfer}, {Laher}, {Mahabal}, {Masci},
  {Nordin}, {Riddle}, {Rusholme}, {Sharma}, {van Santen}, {Shupe}, \&
  {Soumagnac}}]{vanVelzen20}
{van Velzen}, S., {Gezari}, S., {Hammerstein}, E., {et~al.} 2020{\natexlab{b}},
  arXiv e-prints, arXiv:2001.01409

\bibitem[{{Vazdekis} {et~al.}(2015){Vazdekis}, {Coelho}, {Cassisi},
  {Ricciardelli}, {Falc{\'o}n-Barroso}, {S{\'a}nchez-Bl{\'a}zquez}, {La
  Barbera}, {Beasley}, \& {Pietrinferni}}]{Vazdekis15}
{Vazdekis}, A., {Coelho}, P., {Cassisi}, S., {et~al.} 2015, \mnras, 449, 1177

\bibitem[{{Virtanen} {et~al.}(2020){Virtanen}, {Gommers}, {Oliphant},
  {Haberland}, {Reddy}, {Cournapeau}, {Burovski}, {Peterson}, {Weckesser},
  {Bright}, {van der Walt}, {Brett}, {Wilson}, {Jarrod Millman}, {Mayorov},
  {Nelson}, {Jones}, {Kern}, {Larson}, {Carey}, {Polat}, {Feng}, {Moore}, {Vand
  erPlas}, {Laxalde}, {Perktold}, {Cimrman}, {Henriksen}, {Quintero}, {Harris},
  {Archibald}, {Ribeiro}, {Pedregosa}, {van Mulbregt}, \&
  {Contributors}}]{Virtanen_2020}
{Virtanen}, P., {Gommers}, R., {Oliphant}, T.~E., {et~al.} 2020, Nature
  Methods, 17, 261

\bibitem[{{Worthey} \& {Ottaviani}(1997)}]{Worthey97}
{Worthey}, G., \& {Ottaviani}, D.~L. 1997, \apjs, 111, 377

\bibitem[{{Yang} {et~al.}(2008){Yang}, {Zabludoff}, {Zaritsky}, \&
  {Mihos}}]{Yang08}
{Yang}, Y., {Zabludoff}, A.~I., {Zaritsky}, D., \& {Mihos}, J.~C. 2008, \apj,
  688, 945

\end{thebibliography}

\end{document}